# *How can I investigate causal brain networks with iEEG?*

**Yuhao Huang, MD[1] and Corey Keller, MD, PhD[2,3]**

Department of [1]Neurosurgery and [2]Psychiatry & Behavioral Sciences, Stanford University Medical Center, Stanford, CA, 94305, USA

[3]Veterans Affairs Palo Alto Healthcare System, and the Sierra Pacific Mental Illness, Research, Education, and Clinical Center (MIRECC), Palo Alto, CA, 94394, USA

**Abstract.** While many human imaging methodologies probe the structural and functional connectivity of the brain, techniques to investigate cortical networks in a causal and directional manner are critical but limited. The use of iEEG enables several approaches to directly characterize brain regions that are functionally connected and in some cases also establish directionality of these connections. In this chapter we focus on the basis, method and application of the cortico-cortical evoked potential (CCEP), whereby electrical pulses applied to one set of intracranial electrodes yields an electrically-induced brain response at local and remote regions. In this chapter, CCEPs are first contextualized within common brain connectivity methods used to define cortical networks and how CCEP adds unique information. Second, the practical and analytical considerations when using CCEP are discussed. Third, we review the neurophysiology underlying CCEPs and the applications of CCEPs including exploring functional and pathological brain networks and probing brain plasticity. Finally, we end with a discussion of limitations, caveats, and directions to improve CCEP utilization in the future.

## Introduction

The brain connectome, a representation of the functional and structural connection amongst neural elements, has been indispensable for understanding normal and pathological brain activity. Tight interconnections between cortical and regions are known to underlie important motor, perceptual, and cognitive processes. Characterizing the brain connectome involves mapping the neural elements and the inter-regional pathways that connect them, whether directly or indirectly. Several experimental approaches exist to probe the human brain connectome ranging from non-invasive imaging modalities to invasive electrophysiology. In this chapter, we present the notion of cortico-cortical evoked potentials (CCEP) and place it within the context of other brain connectivity measurements.

## Types of Brain Connectivity

The anatomical connections between neural elements define the structural connectivity. At the microscale this constitutes neurons and the synaptic connections between neurons. Attempting to map the complete structural connectivity at this scale might be accomplished in animal models, but is not immediately feasible in humans. Instead, on a macroscale, structural connectivity can be characterized by a set of inter-areal pathways connecting regionally distinct brain regions. Although there is not a unified division of the human brain, numerous parcellation schemes are widely available and define anatomical brain areas at the macroscale [1, 2]. Commonly, diffusion tensor imaging (DTI) and computational tractography map white-matter tracts non-invasively in the human brain. The quantitative outputs from DTI approaches can be tracked longitudinally to assess for changes over time and can be compared between healthy and pathological states [3, 4]. Although the white matter structural connections amongst brain regions both enable and constrain information flow, they do not imply functionality nor directionality. For instance, fiber tracts between two regions may or may not be utilized in the context of certain cognitive states, and the flow of information cannot be established at a particular time based on the sole presence of anatomical links.

In contrast, the correlated neural activity amongst different brain regions defines functional connectivity. Two regions are functionally connected if the neurophysiological activities in those two regions are statistically dependent. This approach enables delineating brain networks in the context of different brain states, and enables characterization of inter-regional communication in a dynamic fashion. Commonly, brain activity on a macroscale is indexed by electrophysiological recordings (electroencephalography) or functional neuroimaging (functional MRI). One emerging method to determine functional connectivity is called cortico-cortical evoked potential (CCEP) mapping and is the focus of this chapter. Unlike other methods of mapping functional connectivity, CCEP is a measure of causal influence between brain regions studied. CCEP mapping can not only provide strong evidence for functional connectivity, but also provide evidence for flow of information, making it a versatile approach to studying brain network dynamics.

**History of CCEP**

CCEPs were first introduced by Matsumoto and colleagues in 2004 when they recruited subjects with intractable epilepsy implanted with invasive subdural electrodes and applied single pulse electrical stimulation [5]. They measured the electrical response to the single pulse at local and remote regions relative to stimulation. These response curves or CCEPs have a characteristic waveform that is dependent on the area being recorded, distance from stimulation site, cortical orientation with respect to gyri/sulci, proximity to white matter tracts, degree of pathology, and brain state. Given that CCEPs have excellent spatiotemporal resolution, do not require participant performance on cognitive tasks, and provide measures of causality and directionality, CCEPs have been used extensively in recent years [5–16].

## Methods and Quantification of CCEP

**Eliciting and Recording CCEPs**

CCEP mapping is an invasive electrophysiological approach relying on electrical measurements from implanted subdural electrodes (electrocorticography or ECoG) or depth electrodes (stereoelectroencephalography or sEEG). In both cases, electrodes are implanted in various brain regions clinically for seizure mapping in patients with epilepsy; however, clinical populations with implanted electrodes are now expanding to those with intractable pain, depression and other neuropsychiatric disorders [17]. To obtain CCEPs, electrical current varying between 1-10mA is delivered to these intracranial electrodes. The current can be delivered in a monopolar or bipolar manner. In monopolar stimulation, the ground electrode can be chosen far away from the stimulation electrode, usually in distant white matter or extracranial space. In bipolar stimulation, the current is delivered between a pair of adjacent electrodes. The current amplitude chosen is dependent on the patient. Typically, less current is needed to activate cortical regions using bipolar compared to monopolar stimulation [18]. Ideally the current amplitude is maximized to improve signal to noise ratio (SNR), but not so high that it would be perceived by the participant (thus direct cortical evoked potentials would be confounded with non-specific sensory evoked potentials) or would induce undesired clinical side effects such as epileptiform discharges. Electrical stimulation pulses result in a local neural response as well as in a distal response. The magnitude of the distal responses is related to the absolute distance and the functional connectivity between the recording and stimulating locations [15]. Stimulation pulses are typically applied between 20 to 150 times at 0.2 to 1Hz frequency [7, 15]. For studies with shorter inter-stimulus interval (ISI) time period, it is important 1) to add a small amount of jitter between pulses to prevent neural entrainment and 2) to ensure that there are no lasting effects such as long-term depression, which often occurs at 1Hz ISI [15, 19]. The number of pulses required for sufficient signal-to-noise (SNR) vary, and in part depends on the quality of the recording amplifier, electrode resistance, and the strength of connection between the stimulating and recording sites.

**Design of CCEP experiments**

There are several experimental considerations when designing a study with CCEP mapping. One common objective is to index the underlying effective connectivity between the site of stimulation and other recording sites [20]. In this case, repetitive pulses of electrical stimulation are delivered at several stimulation sites of interest and CCEPs at all other sites are obtained. CCEP mapping can also be used to index changes in brain connectivity after an experimental intervention. These interventions can be in the form of a behavioral task [13] or a stimulation paradigm designed to alter connectivity on a short timescale [7, 15]. For instance, to determine if effective connectivity is altered after a session of high-frequency stimulation, CCEP pulses can be delivered before and after the intervention to determine the impact on effective connectivity [7].

**Analysis of CCEPs**

Different institutions implement different analytical pipelines to analyze CCEPs. Here we outline a common approach to constructing a CCEP analysis pipeline. First, the local field potential (LFP) signal comprising CCEPs is sampled at ≥500 Hz to allow sufficient sampling of high gamma activity (70-200Hz). Three main steps in CCEP analysis pipelines include pre-processing

(removal of line noise, detrending, and demeaning), artifact removal, and re-referencing. First, line noise can be removed using a bandstop (Notch) filter or a discrete Fourier transform (DFT) filter and data should be detrended and demeaned. Second, electrical artifacts due to stimulation should be removed, especially if power analysis is to be performed. These artifacts are typically present for the first 5 or 10 ms after stimulation, depending on amplifier type, sampling rate, and stimulation amplitude. There are several methods to remove these artifacts including simple interpolation [21], template rejection [22] and replacement with underlying neural data [23]. Third, a re-referencing scheme is selected to remove common noise from the recording channels. A variety of re-referencing schemes exist with advantages and disadvantages associated with each one [24] (see also Chapter 27). For SEEG electrodes, bipolar or laplacian re-referencing is commonly preferred [7]. For subdural electrodes, data are often re-referenced using common average or laplacian methods [7]. After these steps are performed, data are epoched and CCEPs can be quantified using peak analysis in the time domain (N1, N2) or power analysis in the frequency domain (delta 1-4Hz, theta 4-8Hz, alpha 8-12Hz, beta 12-25Hz, gamma 25-70Hz and high-gamma 70-200Hz). In the time domain, the morphology of CCEPs varies significantly and depends on both the stimulation and recording sites. Typically, CCEPs consist of an early potential (termed N1, 10-50ms) and a later potential (termed N2, 50-300ms, [5]) (Figure 1A,B). The amplitudes of the N1 and the N2 potentials can be commonly quantified by averaging the signal, by calculating the peak to trough amplitude or by determining the area under the curve (AUC) within a specified timeframe. Notably, filtered CCEPs, in particular in the high-gamma range, have also been used to quantify effective connectivity [23].

## Applications of CCEPs

**Investigate inter- and intra-regional connectivity of functional brain networks**

CCEP mapping has been used extensively to map the causal, inter-regional connectivity in the human brain. CCEP mapping offers several advantages over other non-invasive mapping tools as it produces *directional* and *causal* measures of connectivity. Luders and colleagues first mapped the language network with CCEPs, reporting reciprocal connections between Broca's and Wernicke's regions [5] as well as within motor cortex [9]. In later work, CCEPs were utilized to demonstrate strong frontal and parietal intralobar connections with associated unidirectional frontal-to-temporal connections with rare temporal-to-frontal connections [10]. CCEPs have also been used to examine connections to and from the hippocampus [6, 14], as well as sensorimotor

[16], visual [13], and recently default mode [25] networks. In summary, CCEP mapping has begun to reveal causal connectivity within and between well-known human functional networks. We predict this work will continue to expand as CCEP mapping becomes commonplace in the epilepsy monitoring unit as a part of routine brain mapping.

**Comparing CCEP mapping to other non-invasive connectivity methods**

A comparison of CCEP mapping to other known non-invasive methodologies that probe anatomical connectivity (diffusion tensor imaging; DTI) and functional connectivity (resting state functional MRI; rs-fMRI) is necessary to 1) identify components of CCEPs that track standard connectivity measures and 2) provide direct electrophysiology grounding to non-invasive connectivity measures that indirectly map neural activity. For example, if the N1 (10-50ms) reflects direct cortico-cortical connectivity, then DTI structural measures should correspond to the N1. Indeed, the number of tracts between two regions positively correlates with the strength of the N1 response and negatively with the latency of the N1 response in the CCEP [8] (Figure 1C). CCEPs have also been compared to rs-fMRI measures, where co-variations of ultraslow (<0.1Hz) fluctuations of the BOLD signal map functional brain networks. However, the neural underpinnings of rs-fMRI are not clear. We hypothesized that fast electrically-propagated potentials observed with CCEP mapping would travel in a similar trajectory as BOLD co-variations. We showed that regions with higher BOLD correlations demonstrated stronger CCEP N1 and N2s. These findings were replicated across patients and functional networks and demonstrated that temporal correlations of slow, spontaneous hemodynamics reflect similar functional interactions to those arising from fast electrically propagated activity [26] (Figure 1D). Specifically, a recent study has shown that the CCEP derived network resembles functional connectivity (as measured by resting state correlation) in channels local to the stimulation site, whereas remote CCEP network correlates best with structural connectivity as measured by DTI [50]. In summary, CCEP mapping provides a direct measure of effective brain connectivity compared to non-invasive brain connectivity measures. The work summarized here demonstrates that the N1 of the CCEP partially reflects structural [8] and functional [26] connectivity while the N2 at least partially reflects functional connectivity [26].

**CCEP mapping to measure pathophysiological networks**

Just as CCEP mapping can probe functional networks, this methodology can also investigate the pathophysiology associated with neurological and psychiatric disorders. As CCEPs are typically performed during intracranial monitoring for seizure localization, epilepsy is by far the most common pathology explored. As seizures arise infrequently, clinicians typically welcome quantitative methods such as CCEP mapping that help explore regions that initiate and propagate seizures. Furthermore, as mounting evidence indicates that seizures arise from a set of interconnected brain regions, tools such as CCEP mapping have become increasingly helpful to delineate these interconnected brain regions that may require surgical resection. Between its potential aid in determining seizure networks and ease of implementation -- CCEP mapping typically takes < 1 hour and does not require patient participation -- CCEP mapping has become increasingly standard in the workup of seizure localization.

Because an abnormal excitation/inhibition balance may indicate regions that can generate seizures ('seizure onset zone'), the CCEP may reflect this imbalance when stimulating (or recording) in the seizure onset zone. Indeed, compared to control regions, larger amplitude N1 was observed following single pulse electrical stimulation to the seizure onset zone [12]. When stimulating the seizure onset zone, CCEP amplitude can also predict the onset of ictal events [27]. In addition to changes in the N1 or N2, 'afterdischarges' have been observed 200-1000ms after electrical stimulation (after the N1 and N2). These later voltage deflections consisting of 'spikes' or 'sharp waves' due to enhanced excitation appear to localize to seizure onset zones [28] and a poorer surgical outcome has been observed when tissue eliciting these afterdischarges were not resected [28]. In summary, both standard CCEP N1 amplitude and afterdischarges are powerful complements to standard methods to localize pathological tissue.

**CCEP mapping to probe brain plasticity**

CCEPs probe effective brain connectivity in a cross-sectional manner. However, CCEPs can also measure changes in cognitive state [13] and probe brain plasticity [7, 15]. To characterize how repetitive stimulation can induce brain plasticity, we applied electrical stimulation in a bipolar fashion patterned to mimic non-invasive transcranial magnetic stimulation (TMS) treatments. Electrical stimulation patterned at 10Hz modulated a subset of brain regions measured by changes in CCEPs assessed pre/post intervention (Figure 3). Modulated regions exhibited stronger baseline CCEPs and could be predicted based on a combination of anatomical (distance from stimulation site) and effective (CCEPs) connectivity. Furthermore, we were able to assess changes occurring *during* 10Hz electrical stimulation using CCEPs elicited from the last pulse in each stimulation train [7]. 'Intratrain' CCEPs – those derived from the last pulse in a stimulation train – changed with subsequent stimulation trains. These intratrain CCEPs also predicted CCEP changes that outlasted the stimulation intervention. Together, this work demonstrated that CCEPs can be used to probe brain plasticity and that baseline connectivity profiles can be used to predict regions susceptible to stimulation-induced brain changes. Future human plasticity studies will focus on how CCEPs are modulated as a function of stimulation frequency, duration, amplitude, and brain state.

## Mechanistic basis of CCEPs

The neurophysiological mechanisms underlying CCEPs are only partially known and future work to further elucidate these mechanisms will improve insights from CCEP studies. Here, we provide an overview of what is known about the mechanistic basis of CCEPs. Although CCEPs are applied using different stimulation parameters, the most common form is with *bipolar, biphasic stimulation*. Biphasic stimulation allows for a balanced charge to be delivered and bipolar stimulation provides a more local stimulation of cortex compared to monophasic stimulation [29], thus minimizing the spatial spread of electrical stimulation.

**Neurophysiology at site of stimulation**

Electrical stimulation triggers multiple local cortical events that determine if pyramidal neurons fire and action potentials propagate to remote regions. Initial electrical stimulation depolarizes superficial dendrites of layer V pyramidal neurons and layer II/III inhibitory neurons that synapse on layer V neurons. The balance of these two events determine if layer V pyramidal neurons will fire; that is, the firing of many GABAergic neurons will hyperpolarize layer V neurons and inhibit their firing and orthodromic propagation of action potentials, whereas sufficient pyramidal dendritic depolarization without GABAergic activation will lead to pyramidal neuron firing and orthodromic propagation. In addition to this neuronal interplay, electrical stimulation will also directly depolarize long-range axons traversing the stimulated region and generate action potentials propagating orthodromically to distant synapses as well as antidromically backpropagating to pyramidal cell soma [30, 31]. In summary, electrical stimulation activates white matter tracts both physiologically through pyramidal cell firing from dendritic depolarization and non-physiologically through direct depolarization of traversing white matter tracts.

**Potential propagation pathways**

Evidence from animal studies have shown that 1) direct cortical stimulation elicits superficial-to-deep propagation of cortical layers [1][2]locally at stimulated cortex [32] and 2) orthodromic axonal activation occurs far more than antidromic activation [33]. Data from human studies support the fact that electrical stimulation primarily activates neurons in middle layers [34] and that middle layer pyramidal neurons typically propagate to mono- and poly-synaptically connected regions via cortico-cortical and cortico-subcortical-cortical projections [32] (Figure 2). Together this work suggests that CCEPs propagate via both a major pyramidal cell contribution via orthodromic cortico-cortical and cortico-subcortical-cortical projections as well as a minor antidromic contribution [9, 34].

**Electrophysiology underlying the N1 and N2 of the CCEP**

Typical CCEP patterns consist of an early N1 peak within the first 50ms and a later N2 slow wave lasting up to 500ms [9, 20, 34]. Direct electrical stimulation has been shown to activate pyramidal neurons mono-synaptically connected to the site of stimulation within 4-8ms of stimulation [35, 36]. Unfortunately, clinical grade amplifiers saturate for up to 10ms, masking this initial response. Instead, the N1 observed after 10ms in clinical amplifiers likely reflect oligo- and poly-synaptic pyramidal responses local or in close proximity to the applied stimulation. Several lines of evidence are consistent with this notion, including 1) excitatory neuronal responses within 50ms (N1) after direct electrical stimulation of cat cortex [37]; 2) neuronal spiking within 50ms (N1) after direct electrical stimulation in human cortex [38]; 3) increases in multi-unit activity in deep (layer IV-VI) cortical layers after single pulse stimulation in humans [34]. In contrast, the N2 slow wave in humans is time-locked to a reduction in both spiking [38] and middle-to-deep layer multi-unit activity [34], likely reflecting a prolonged inhibitory time period. In summary, single and multi-unit recordings of the responses to cortical stimulation suggest that the N1 evoked response reflects early pyramidal neuron activation while the N2 response represents long-lasting inhibition.

It is worth noting that this pattern of brief excitation followed by long-lasting inhibition is not specific to CCEPs and found in other physiological and pathophysiological neuronal events. These include the neural response to epileptic discharges [38, 39] and single pulses of TMS [40]. As such, studying these events will yield further insight into the neurophysiological basis of this excitation-inhibition wave and these processes can be evoked and perturbed.

## Advanced considerations

**Limitations and caveats**

Although CCEP mapping represents an important method to causally and directionally probe the human brain, several limitations are worth discussing. First, as mentioned above, because common clinical amplifiers saturate during the first 10ms after electrical stimulation, important mono- and di-synaptic connections may be masked by simulation artifacts. As such, novel methods to remove electrical stimulation artifacts analytically [23, 41] as well as the transition to research grade stop-and-hold amplifiers or those with high dynamic range are currently underway. These modifications will be critical to elucidate cortical responses within the first 10ms. Second, without a clear understanding of the neural basis of the strength, latency, and polarity of the N1 and N2, as well as the degree that CCEPs reflect orthodromic vs antidromic propagation patterns, it is difficult to accurately interpret results from CCEP studies. As underlying neurobiological mechanisms are elucidated, more accurate interpretation of the CCEP will be possible. Third, the variability of experimental design across institutions and even across patients within an institution limits the ability to compare across studies. These parameters include electrode type (stereo-EEG vs. subdural implantation), stimulation configuration (monopolar, bipolar), current amplitude (from 1-10mA), pulse duration (100-500µs), inter-stimulation interval (0.5-10s), and number of stimulation repetitions (10-500). Other parameters that are more difficult to control and often not reported include the patient's cognitive state, type and dosage of anti-epileptic medication, time of day, and level of sleep deprivation. A thorough reporting of each of these parameters in a reproducible manner is critical to assess differences across centers and studies. Fourth, a common criticism in CCEP studies (and all intracranial studies in epilepsy patients) is the lack of generalizability of results (see also Chapter 6). However, several lines of evidence can minimize this concern: 1) as these patients vary with respect to seizure semiology and localization, consistent results across patients are not likely due to the patient's pathophysiology; 2) electrodes involved in the seizure onset zone and early seizure spread are often removed prior to analysis so that results are from non-pathological regions.

## Future CCEP mapping approaches

In the future, three key improvements will be critical for the CCEP field. Over the past five years, an increasing number of hospitals have incorporated CCEP mapping as a standard approach for patients with implanted electrodes. As the community increases, more sophisticated quantitative methods will be applied to analyze CCEP data. The first key advance, as discussed before, is the transition to research grade amplifiers to observe neural information within 10ms of the electrical pulse and potentially study mono- and di-synaptic connections. A second key advance comes from the improvement in computational power over the past decade. A typical CCEP study involves analysis of the N1 and N2 individually at each recording electrode for each stimulation site. Recent increases in computational power will enable data driven multivariate approaches by considering the N1 and N2 timing and amplitude *between* recording electrodes and stimulation sites. Examples include graph theory frequency decomposition and multivariate machine learning approaches. A typical CCEP study uses *divergent* analysis to examine CCEPs by comparing CCEP waveforms at multiple electrodes after stimulation of one region. In contrast, a recent study has explored the potential use of *convergent* CCEP analysis, where CCEP responses at the same electrode are compared after the sequential stimulation of multiple other regions, with sufficient time between stimulation of each region (Figure 4). Convergent analysis controls for the variability of electrode orientation, gray/white matter, etc in interpreting the response [42]. Another computationally-intensive method explores the use of *basis curves* to more accurately define the CCEP. Although the N1 and N2 are useful ways to describe time windows around the CCEP, in reality there is significant variability with respect to timing and polarity of each potential, even within a given patient. As such, Miller and colleagues recently used computationally-driven tools such as basis curves to extract the principal components of CCEP waveforms [51] (Figure 4). Using basis curves to explore how these more fundamental components contribute to physiology and pathology across brain regions will open up new avenues of research. A third critical improvement comes from open science and big data initiatives [43]. Here, CCEP data format standardization will enable big data computational approaches not possible before (see also Chapter 45). For example the Functional Brain Tractography database (f-tract.eu) is a dataset with CCEP recordings from >1500 patients with standardized data formatting [44]. Together, these improvements will pay the way for the next generation of CCEP mapping: standardized, data-driven, and reproducible.

Several other research avenues for CCEPs should be explored. First, CCEPs are applied using different stimulation parameters across groups and it will be important to explore how the stimulation parameter space differentially affects brain dynamics. Second, recent studies have explored how brain state influences CCEPs, either using cognitive tasks [13] or awake/sleep patterns [45]. Future work will further explore how cognitive states differentially influence CCEP amplitude, latency, and propagation patterns. Third, the combination of intracranial and non-invasive stimulation and recording methods will be critical to bridge the knowledge gap that exists between intracranial and non-invasive brain recordings and move towards more generalizable tools that can be utilized in outpatient clinics. Recent combined recordings have mapped the relationship between CCEPs measured both intracranially and non-invasively [46]. Fourth, implantation of deep electrodes for the monitoring, understanding, and treatment of neuropsychiatric illness is increasingly being performed [17, 47–49]. For instance, depth electrode implantations are being done for patients with obsessive compulsive disorder (OCD) [52] or major depressive disorder (MDD) [53]. In both cases, depth electrodes implanted in multiple regions were used to assay network responses to tasks and to mood variations. Hence, one can expect that as electrodes are implanted for MDD, OCD, obesity, pain, and many other disorders, CCEP mapping will become a mainstream method of evaluation of cortical excitability and plasticity.

## Conclusion

CCEP mapping represents an important methodological tool to quantify brain connectivity in humans. CCEP mapping has excellent spatial and temporal resolution with *electrophysiological grounding* as with any intracranial approach with the additional characteristics of *causality* and *directionality*. The N1 of the CCEP likely reflects early excitation of pyramidal neurons from feedforward connections while the N2 likely represents a long-lasting inhibitory period from feedback connections. The N1 reflects at least in part structural connectivity between two regions while the N2 can be influenced by brain state and cognitive demands. Future improvements in amplifier design, computational power, and big data sharing initiatives, as well as the expansion of implanted electrode approaches to indications other than epilepsy will pave the way for the next generation of CCEP mapping. This underutilized tool in basic and clinical neuroscience represents a powerful tool to investigate brain connectivity, the causal involvement of brain regions in cognitive processes, and the pathological networks in neurological and psychiatric disorders.

**Figures and Legends**

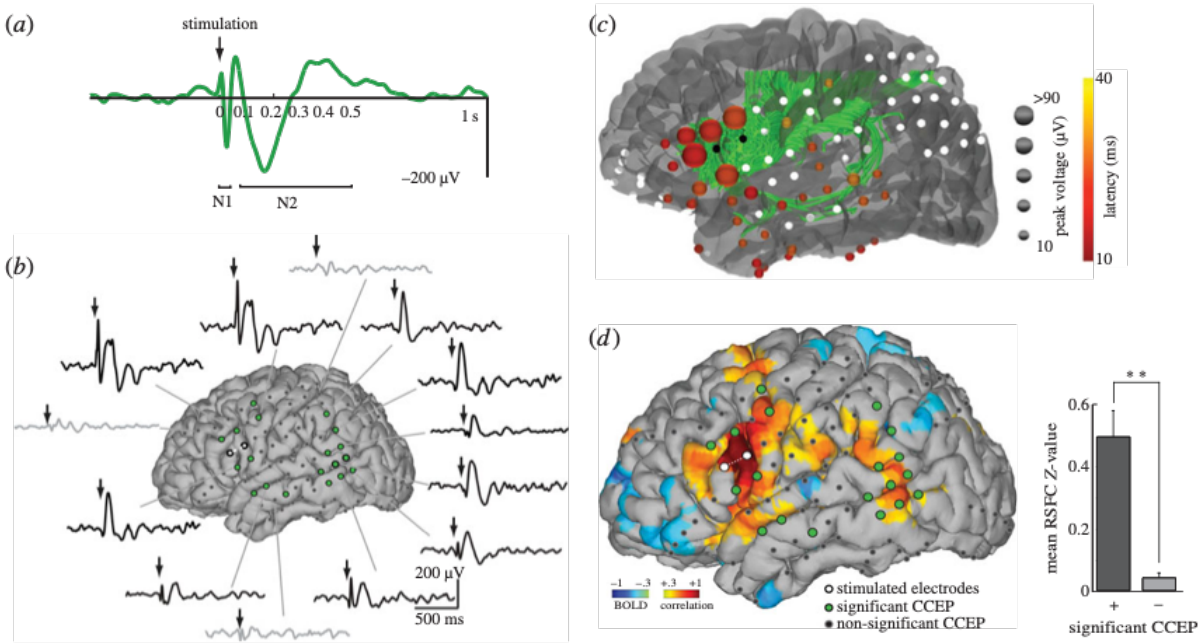

*Figure 1 - CCEPs and their relationship to anatomical and functional connectivity.* A) CCEPs typically consist of an early N1 (10-50ms) and a later N2 (50-250ms) response. B) Example of CCEP maps (electrical stimulation to white electrodes). C) Comparison of CCEP (effective connectivity) and DTI (structural connectivity). Here, the number of white matter tracts measured with DTI are positively correlated with the strength of the CCEP N1 and negatively correlated with its latency. Adapted with permission from [8]. D) Comparison of effective and functional connectivity. Here regions exhibiting strong CCEP N1 and N2 also exhibit strong fMRI connectivity. Adapted with permission from [26].

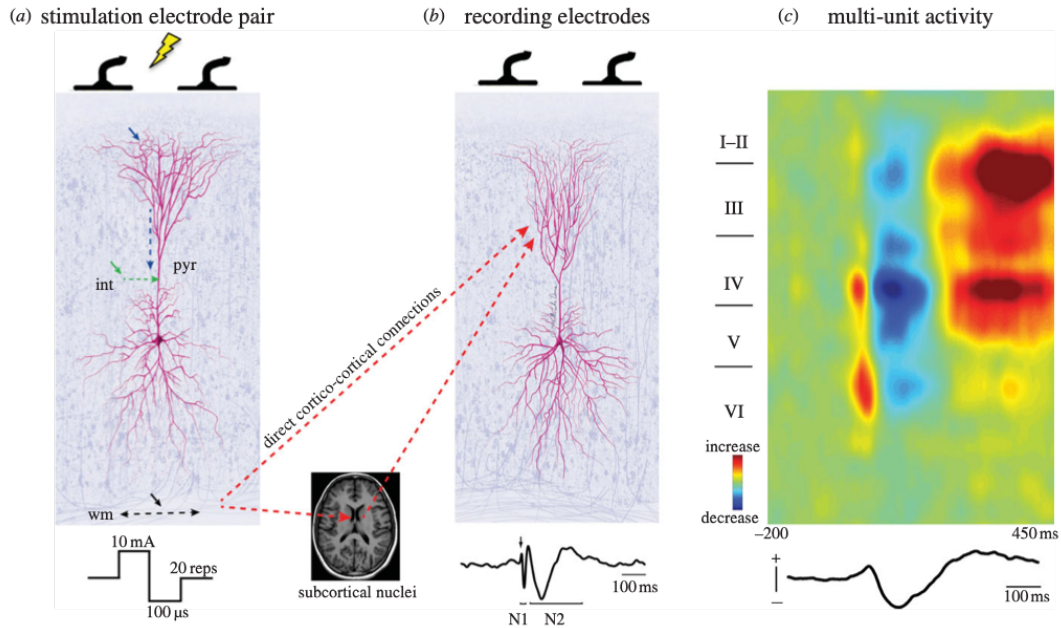

***Figure 2 - Proposed mechanism of CCEP generation.*** A) Electrical stimulation propagates through the cortex to local pyramidal cells via direct dendritic activity (blue arrows), adjacent interneurons (green arrows), or white matter traversing the stimulated region (black arrows). B) Electrical activity from stimulation is propagated to downstream regions via direct cortico-cortical connections and indirect cortico-subcortico-cortical projections. C) Multiunit response to electrical stimulation. CCEP below was derived from a recording from deep layers. The initial N1 response is accompanied by an increase in mid-deep increase in multiunit activity, while the later N2 is accompanied by a suppression/activation pattern. Adapted with permission from [34].

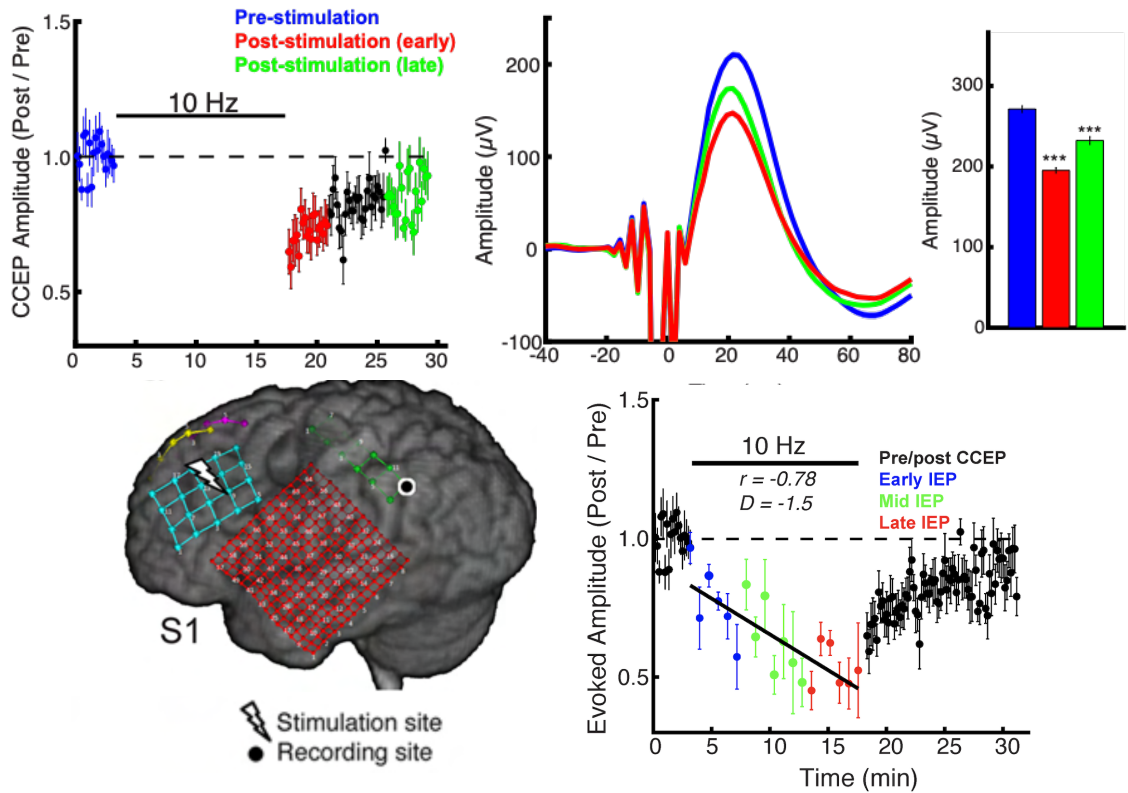

**Figure 3 - Repetitive electrical stimulation modulates the CCEP.** In this representative stimulation-recording pair in one subject, electrical stimulation was applied to the prefrontal cortex (lightning bolt) and recordings were measured in the parietal cortex. Repetitive 10Hz electrical stimulation (5s on, 10s off, 3000 total pulses) modulated the CCEP for ~20 minutes. Top: quantification of CCEP amplitude over time. Bottom: by evaluating the CCEP after the last pulse of each 10Hz stimulation, one can evaluate the evolution of plasticity effects induced by repetitive stimulation or other interventions. Adapted with permission from [7] and [15].

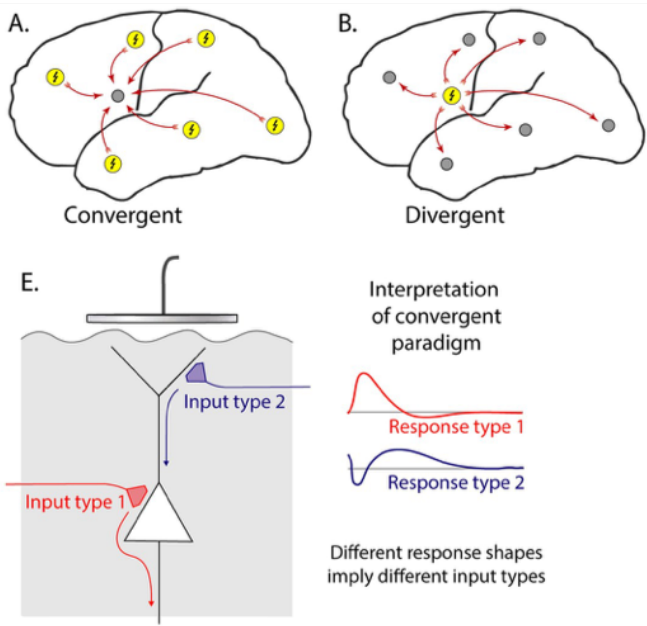

**Figure 4 - Convergent and divergent CCEP analysis paradigms.** A) *Convergent* - CCEPs at one region (gray circle) are compared with the effect of stimulating all other regions (yellow circle). B) Divergent - The CCEP at all regions are examined and compared in response to stimulation of a chosen region. Adapted with permission from [51].